\documentstyle[psfig,epsfig,12pt]{article}
\begin{document}
\baselineskip 0.25in
\begin{center}
\large {\bf Money Exchange Model and a general outlook} 
\end{center}
\smallskip
\begin{center}
Abhijit Kar Gupta
\end{center}

\begin{center}
{\it Physics Department, Panskura Banamali College\\
Panskura R.S., East Midnapore, WB, India, Pin-721 152}\\
{\em e-mail:}~{abhijit$_{-}$kargupta@rediffmail.com}
\end{center}

\abstract{The Kinetic Gas theory like two-agent money exchange model, recently introduced 
in the Econophysics of wealth distributions, is revisited. The emergence of Boltzmann-Gibbs 
like distribution of money to Pareto's law in the tail of the distribution is examined in 
terms of $2\times 2$ transition matrix with a general and simplified outlook. Some additional 
interesting results are also reported.}

\smallskip

\noindent{\bf Introduction}

\smallskip

{\it Econophysics of Wealth distributions} \cite{eco} is an emerging area where some Statistical 
Physicists and Economists have been engaged in interpreting real economic data of money, wealth 
and income of all kinds of people pertaining to different societies and nations. Economic activities 
have been assumed to be analogous to elastic scattering processes \cite{chak1, chak2, yako}. 
Analogy is drawn between Money ($m$) and Energy ($E$) where temperature ($T$) is average money ($<m>$) 
of any individual at equilibrium.
Some early attempts \cite{early} have been made to understand the income distributions which follow Pareto's law
($P(m) \propto 1/m^{\alpha+1}$) at the tail of the distributions with the index ($\alpha$) varying from 1 to 2.5. 

Kinetic Gas theory like models are introduced to exploit apparent similarities between a many particle 
system and a social system of many agents. The primary attempt is to understand the distributions of 
money/ income corresponding to different classes of people. 
We revisit such a Kinetic theory like two-agent money exchange model as recently proposed by Chakrabarti 
and group \cite{chak2, chak3}.
In this model any two agents chosen randomly from a number of agents ($N$) are allowed to interact (trade) 
stochastically and thus money is exchanged. Stochasticity is 
introduced in terms of a parameter $0<\epsilon<1$ into the interaction. One arrives at a
 Boltzmann-Gibbs (exponential)-type distribution ($P(m) \propto {\exp(-\beta m)}$) of 
individual money. In the next stage, a saving propensity factor ($\lambda$) is incorporated to show that 
the distribution shifts away from the exponential distribution. A peak appears at a value other than 
at zero. In the later stage the saving propensity factor is made random (among the agents) but 
frozen in time. This brings a distribution qualitatively different from the earlier: one gets a power 
law at the tail of the distribution. This indicates the emergence of Pareto's Law ($P(m) \propto 1/m^{\alpha +1}$) at the tail of the distribution which is well known among Economists/ Econophysicists. 
In the above class of models total money ($M=\sum_im_i$) of all the agents is invariant in time. Also the money ($m$) is 
conserved locally which means the sum of money of two agents before and after trade (interaction) 
remains constant: $m_i(t+1)+m_j(t+1)=m_i(t)+m_j(t)$. 

We systematically examine the role played by the parameters $\epsilon$ and $\lambda$ in 
this conserved class of models and attempt to understand the emergence of behaviours in terms of 
probability distribution functions ($P(m)$) of money ($m$). Also we come across additional interesting features in the model 
with variable saving propensity ($\lambda$). 

The time evolution of money in the model without saving propensity is described as follows:
\begin{equation}
m_i(t+1)=\epsilon (m_i(t)+m_j(t)) 
\end{equation}

\begin{equation}
m_j(t+1)=(1-\epsilon)(m_i(t)+m_j(t)),
\end{equation}

Therefore, we can say that the distribution in money is evolved through $2\times 2$ transition matrices ($T$):
\[T=\left(\begin{array}{cc}
\epsilon & \epsilon\\
1-\epsilon & 1-\epsilon
\end{array}\right)\]

The above matrix is, however, singular which means the inverse of this matrix does not exit. 
This indicates that the evolution through such transtion matrices is bound to be irreversible. 
As a result we get exponential (Boltzmann) distribution of money. This can be perceived here in a 
different way too. When we take a product of such matrices and we get back one which is nothing but 
the leftmost matrix: 
\[\left(\begin{array}{cc}
\epsilon & \epsilon\\
1-\epsilon & 1-\epsilon
\end{array}\right)\left(\begin{array}{cc}
\epsilon_1 & \epsilon_1\\
1-\epsilon_1 & 1-\epsilon_1
\end{array}\right)=\left(\begin{array}{cc}
\epsilon & \epsilon\\
1-\epsilon & 1-\epsilon
\end{array}\right)\]

The above signifies the fact that when two agents happen to interact repeatedly (via this kind of 
transition matrices), the last of the interactions is what matters (the last matrix of the product survives). 
This 'loss of memory' may be attributed to the path to irreversibility in time here. 

Let us play the game in a different way. Suppose two agents trade (interact) in such a manner that 
we may arrive at the following general transition matrix:  
\[T_1=\left(\begin{array}{cc}
\epsilon_1 & \epsilon_2\\
1-\epsilon_1 & 1-\epsilon_2
\end{array}\right)\]

\noindent where $\epsilon_1$ and $\epsilon_2$ are two independent random fractions varying between 0 and 1 
(but not equal). This simply signifies that $\epsilon_1$ fraction of money of the 1st agent added 
with $\epsilon_2$ fraction of money of the 2nd agent is retained by the 1st agent after the trade.
The rest of their total money is shared by the 2nd agent. This is a little generalization of the 
earlier case. 
However, this matrix is now nonsingular (as long as $\epsilon_1\neq\epsilon_2$) and the two-agent 
interaction process may be said to be reversible in time. Therefore, 
we expect to have an equilibrium distribution of money which may be qualitatively different from the 
earlier exponential (Boltzmann-Gibbs like) one. 
In the above general matrix, if we put $\epsilon_1=1$ and $\epsilon_2=0$, this reduces to an 
Identity matrix $I=(\begin{array}{cc} 1 & 0\\ 0 & 1 \end{array})$
which is the stationary and trivial case (no interaction).

In Fig.1 we plot two distributions (for $\epsilon_1=\epsilon_2$ and 
$\epsilon_1 \neq \epsilon_2$) in support of the above discussions.
The computer simulation is performed on a system of $N$=1000 agents. Two agents are selected randomly 
to interact (trade). No qualitative change is seen in the
distributions when we take different system sizes. A single interaction between any two agents 
is defined here as a single time step. Simulation is done for $10^5$ time steps and averaging
is done over $10^4$ initial configurations. The distributions, we plot here, are not normalized.  

\begin{figure}[htb]
\centerline{\psfig{figure=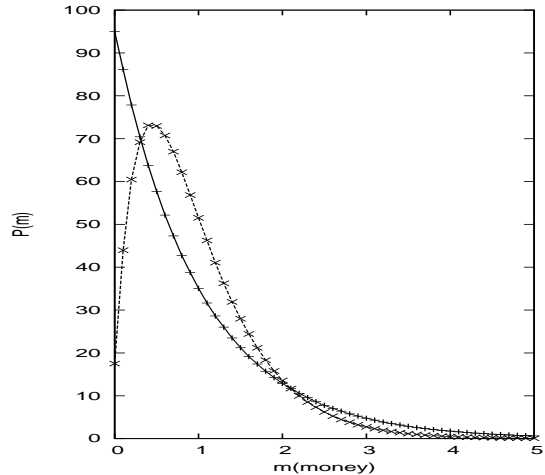, width=3in, height=2.5in}}
\caption{Distribution of money($m$) in two cases: Exponential curve is for
the case when the transition matrix is singular (no time reversal symmetry), the other curve
with a peak is for the case when the transition matrix is made non-singular by choosing 
$\epsilon_1\neq\epsilon_2$.} 
\end{figure}

Let us now look at the case where a saving propensity factor $\lambda$ is incorporated \cite{chak2}. 
When two traders/ agents meet, each of them is supposed to keep aside $\lambda$-fraction (fixed) of 
their individual money (This is like a gamble with precaution.). 
Rest of the money ($(1-\lambda)$-fraction of the sum of their total money) is redistributed (between them) 
with the stochasticity factor $\epsilon$ ($0<\epsilon<1$).
The transition matrix now looks like: 
\[\left(\begin{array}{cc}
\lambda+\epsilon(1-\lambda) & \epsilon(1-\lambda)\\
(1-\epsilon)(1-\lambda) & \lambda+(1-\epsilon)(1-\lambda)
\end{array}\right)\]

We may now rescale the matrix elements by assuming $\tilde\epsilon_1=\lambda+\epsilon(1-\lambda)$ and 
$\tilde\epsilon_2=\epsilon(1-\lambda)$ in the above matrix. Therefore, the above transition matrix reduces 
to
\[T_2=\left(\begin{array}{cc}
\tilde\epsilon_1 & \tilde\epsilon_2\\
1-\tilde\epsilon_1 & 1-\tilde\epsilon_2
\end{array}\right)\]

Thus the matrix $T_2$ is of the same form as $T_1$. Here, as $0<\epsilon<1$ and $\lambda$ is something
between 0 and 1,
we do have $0<\tilde\epsilon_1<1$ and $0<\tilde\epsilon_2<1$. As long as $\tilde\epsilon_1$ and 
$\tilde\epsilon_2$ are dfferent, the determinant ($\Delta=\tilde\epsilon_1-\tilde\epsilon_2=\lambda$) 
of the matrix is nonzero. Therefore, the effect of the saving propensity factor $\lambda$, thus introduced,  
essentially leads to have {\em non-singular} transition matrix.
Hence it is clear from the above discussion that the distribution (in money) would 
likely to be qualitatively no different from what can be achieved with 
transition matrices like $T_1$ with $\epsilon_1$ and $\epsilon_2$ ($\epsilon_1\neq\epsilon_2$), in general.
The distributions obtained for different $\lambda$ 
(as reported in \cite{chak3}) may correspond to the difference in $\epsilon_1$ and $\epsilon_2$ as we see.  

In the next stage, when the saving propensity factor $\lambda$ is made random, the transition matrix 
between any two agents having different $\lambda$'s (say, $\lambda_1$ and $\lambda_2$) would now 
look like: 
\[\left(\begin{array}{cc}
\lambda_1+\epsilon(1-\lambda_1) & \epsilon(1-\lambda_2)\\
(1-\epsilon)(1-\lambda_1) & \lambda_2+(1-\epsilon)(1-\lambda_2)
\end{array}\right)\]
Again we rescale the elements by putting $\tilde\epsilon_1=\lambda_1+\epsilon(1-\lambda_1)$ and 
$\tilde\epsilon_2=\epsilon(1-\lambda_2)$. Hence this matrix can also be reduced to the same form as 
that of $T_2$. 
\[T_3=\left(\begin{array}{cc}
\tilde\epsilon_1 & \tilde\epsilon_2\\
1-\tilde\epsilon_1 & 1-\tilde\epsilon_2
\end{array}\right)\]
The determinant here is $\Delta=\tilde\epsilon_1-\tilde\epsilon_2=\lambda_1(1-\epsilon)+\epsilon\lambda_2$. 
Here also $\Delta$ is ensured to be nonzero as all the parameters $\epsilon$, $\lambda_1$ and $\lambda_2$
have the same range: between 0 and 1. This means that each transition matrix for two-agent money 
exchange remains non-singular which signifies the interaction process to be reversible in time. 
We may also check here that $0<\tilde\epsilon_1<1$ and $0<\tilde\epsilon_2<1$, which again corresponds to 
the transition matrix $T_1$ as discussed before.
Therefore, it may be apparent that 
{\it qualitatively different distributions are possible when we appropriately tune the two 
elements $\epsilon_1$ and $\epsilon_2$ in the general form of transition matrix} $T_1$ (We have 
not done so here in this paper to explicitly demonstrate that further.). 
Nevertheless the incorporation of the 
parameter $\lambda$ goes a step closer to interpret real economic data. 
However, the emergence of power law tail ({\it Pareto's law}) in the distribution 
is not well understood in the model\cite{chak3} we are discussing. 

In this context, we present an additional interesting feature of the model (as proposed in \cite{chak3}) 
with variable saving propensity
($\lambda$). Suppose, we consider $\lambda$ to have only two fixed values $\lambda_1$ and $\lambda_2$ and 
that they are widely different. This may be thought of the society to have only two kinds of people: some of them 
do save a very large fraction (fixed) of their money and the other kind of people who save a very small 
fraction (fixed) of their money. Introducing this binary-$\lambda$ in the model brings out an interesting 
feature: a {\it double-peak distribution} in money which can be seen from Fig.2 (This result is also 
obtained by simulating a system of $N$=1000 agents.). Thus two distinct economic classes appear out of this.
It has been observed that the system evolves towards a distinct two-peak distribution as the difference in 
$\lambda_1$ and $\lambda_2$ is increased systematically. Later it is seen that we still get two-peak 
distributions even when $\lambda_1$ and $\lambda_2$ (one high and the other low) are distributed in a 
narrow range around their fixed values. 

\begin{figure}[htb]
\centerline{\psfig{figure=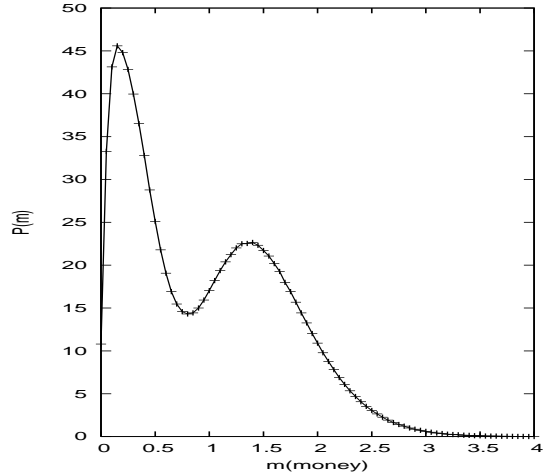, width=3in, height=2.5in}}
\caption{Double-peak distribution of money($m$) with fixed values, $\lambda_1$=0.2 and $\lambda_2$=0.8: 
Emergence of two Economic classes.}
\end{figure}

The class of models here we discuss is conserved. The sum of money of all the agents $M=\sum_i m_i$ is
fixed for all time. This is ensured by the two-body interaction process (rule of the game) where the sum 
of money of the two agents is
conserved before and after interaction as mentiond in the beginning. So the
sum of two elements of a column of a transition matrix 
$T=(\begin{array}{cc}
t_{11} & t_{12}\\
t_{21} & t_{22}
\end{array})$
has to be unity by design: $t_{11}+t_{21}=1$, $t_{12}+t_{22}=1$.  Whatever extra parameter we add in the 
model, no matter, the matrix has to retain this property. (However, an extra parameter, like $\lambda$, 
may help explaining things better.)
Therefore, this kind of conserved models, 
in general, may be understood in terms of a general transition matrix like $T_1$ as discussed in the 
beginning.
$T_1$ essentially tells that one agent retains randomly $\epsilon_1$-fraction ($0<\epsilon_1<1$) of 
his own money added with random $\epsilon_2$-fraction ($0<\epsilon_2<1$) of the other, where the other 
keeps the rest.

\vskip 0.1in
\noindent{\bf Acknowledgment:}

\smallskip

The author is grateful to D. Stauffer for some important comments on the content and style 
of the manuscript. B.K. Chakrabarti and A. Chatterjee are duly acknowledged for the discussions 
on their model.

\end{document}